# Accessing photon bunching with photon number resolving multi-pixel detector

**Dmitry A. Kalashnikov[1], Si Hui Tan[1], Maria V. Chekhova[2,3], and Leonid A. Krivitsky[1,*]**

[1]*Data Storage Institute, Agency for Science Technology and Research, 117608 Singapore*
[2] *Max-Planck Institute for the Science of Light, 91058 Erlangen, Germany*
[3] *M. V. Lomonosov Moscow State University, 119991 Moscow, Russia*
[*]*Leonid_Krivitskiy@dsi.a-star.edu.sg*

**Abstract:** In quantum optics and its applications, there is an urgent demand for photon-number resolving detectors. Recently, there appeared multi-pixel counters (MPPC) that are able to distinguish between 1,2,..10 photons. At the same time, strong coupling between different pixels (cross-talk) hinders their photon-number resolution. In this work, we suggest a method for `filtering out' the cross-talk effect in the measurement of intensity correlation functions. The developed approach can be expanded to the analysis of higher-order intensity correlations by using just a single MPPC.

## 1. Introduction

In modern quantum optics there is a considerable interest in multiphoton quantum states, as they are believed to be beneficial in several practical and fundamental tasks. In particular, in linear optics quantum computing (LOQC), exploiting multiphoton states would enable a step from two-qubit to multiphoton gates, thereby increasing the performance and flexibility of LOQC protocols [1-3]. A special class of multiphoton entangled states, referred to as NOON-states, is believed to find broad applications ranging from the studies of the foundations of quantum mechanics to quantum lithography beyond the diffraction limit [4,5]. Moreover, accurate estimation of the multiphoton component is of crucial importance in the security analysis of single-photon quantum key distribution protocols.

Any practical implementation of multiphoton states would require solving two basic problems, namely (a) efficient generation of such states and (b) their accurate characterization. The first goal is achieved by using high-power lasers in various nonlinear optical processes, where the interaction strength can be significantly enhanced by utilizing tailored nonlinear media (e.g. periodically poled crystals or nonlinear optical waveguides), possibly in conjunction with optical cavities [6]; the second goal, however, represents a considerable difficulty due to the existing limitation in the response of single-photon detectors. Indeed, a conventional avalanche photodiode (APD), operating in the Geiger mode, discriminates only between "zero photons" and "one photon or more" without further photon-number resolution [7]. The urgent requirement to access photon statistics of faint light pulses motivates the investment of considerable efforts in the engineering of photon-number resolving detectors (PNRD) – a class of devices where the produced outcome depends on the number of simultaneously impinging photons.

Up to date, several technologies have been suggested to realize PNRD [8]. First, we mention a family of cryogenic devices such as visible-light photon counters (VLPC) and transition edge sensors (TES) [9-14]. These devices have moderate photon number resolution, low noise, and high quantum efficiency. However, the main limitation of such technology is the requirement of cryogenic cooling (down to 0.1K for TES and 6K for VLPC), which demands for highly skilled operation. An alternative approach for the implementation of PNRD, which does not involve cryogenic cooling, is based on various modifications of widely accessible photomultiplier tubes (PMT) and avalanche photodiodes (APD). A dedicated class of hybrid PMT detectors demonstrates discrimination of up to 3 photons [15]. A resolution up to 5 photons was demonstrated with a single InGAs detector by careful characterization of the early stage of the avalanche [16]. Other alternatives rely on joint measurements by independent APDs, when an incoming pulse is split into different spatial or temporal modes by a chain of beam splitters [17-19]. Though a remarkable resolution of up to 9 photons was demonstrated with the temporal binning, future expansion of such schemes is negatively affected by the necessity to use more beamsplitters and photodetectors.

Recently, a compact and cost-efficient solution for PNRD became commercially available with the introduction of a Multi Pixel Photon Counter (MPPC) [20]. In MPPC several hundred of APDs, referred to as *pixels*, are embedded in a single chip of several millimeter size with their outputs connected into a summation circuit. The chip is illuminated by a spatially diffused light spot (e.g. originating from the fiber), providing that the chance of two photons to hit the same pixel is negligible. The amplitude of MPPC output is proportional to the number of firing pixels, which, in an ideal case, is equivalent to the number of registered incident photons. Thus, the concept of MPPC resembles the traditional approach of separating an incoming pulse into multiple spatial modes, providing, however, a striking advantage in compactness and photon-number resolution.

The MPPC operation was theoretically modeled and experimentally tested in recent works on the reconstruction of photon statistics of thermal and coherent fields [21-25]. Despite the significant benefits outlined above, limitations of MPPC technology have been noted, which hinder its wide applications in quantum optics environment. The most crucial and very unique imperfection of MPPC is the crosstalk between pixels [24]. The crosstalk appears due to poor optical isolation of pixels; as a result, a photon impinging on a given pixel not only triggers an avalanche but also creates a photon, which, in turn, can trigger a simultaneous "fake" avalanche in a neighboring pixel. Since "fake" avalanches cannot be distinguished from the "true" ones, the accurate reconstruction of photon statistics becomes quite complicated.

The influence of the crosstalk is especially evident when one deals with weak states of light with the average number of photons less than one. In this case, some single-photon detection events would be accompanied by "fake" avalanches, thereby imposing additional photon correlations to the original statistics. From a practical aspect it is of interest to analyze two interconnected matters: (a) how do "fake" photon correlations behave for a weak state and (b) whether an existing MPPC allows one to distinguish "true" photon bunching, for instance, in the case of squeezed vacuum.

In the present paper we estimate the potential of the MPPC in the accurate measurements of non-classical photon correlations. We develop a method of inferring the second-order correlation function ($g^{(2)}$) from MPPC data based on a model of MPPC crosstalk. In order to check the consistency of the method, the experimental results, obtained by MPPC, are compared with the results of a traditional Hanbury Brown and Twiss (HBT) experiment. The way of expanding the present analytical formula for the measurement of higher-order intensity correlation functions is also presented. To the best of our knowledge, this is the first attempt to use MPPC in the study of non-classical light states.

**2. Theory**

Characterization of various states of light in quantum optics is conveniently performed in terms of normalized Glauber's correlation functions (CFs). The *l*-th order CF at zero time delay is defined as

$$g^{(l)} = \frac{\langle a^{+l} a^l \rangle}{\langle a^+ a \rangle^l}, \quad (1)$$

where $a^+, a$ are photon creation and annihilation operators, respectively. In this paper we focus on the analysis of the second-order CF, $l = 2$.

As an example, consider normalized Glauber's CFs for two types of states, coherent one and squeezed vacuum one. *Two-mode squeezed vacuum* state can be produced via nondegenerate parametric down conversion (PDC) in a nonlinear medium. In this case the joint state vector is a superposition of Fock-state products with the same photon numbers for signal *(s)* and idler *(i)* beams [26],

$$|\psi_{s,i}\rangle = \sum_{n=0}^{\infty} C_n |n\rangle_s |n\rangle_i \quad (2a)$$

In the case of degenerate PDC, the signal and idler photons are indistinguishable, and the state produced is *single-mode squeezed vacuum*, with the state vector given by a superposition of even-number Fock states,

$$|\psi\rangle = \sum_{n=0}^{\infty} C_{2n} |2n\rangle \quad (2b)$$

The index $n$ does not have to be the photon number for single modes, but may relate to the ensembles of many modes, accessed via multimode detection [26]. In this case, the probability amplitudes obey the Poissonian distribution, $|C_n| = \sqrt{e^{-\langle n \rangle} \langle n \rangle^n / n!}$, and $\langle n \rangle$ is the mean photon number for the whole ensemble (supermode).

The state vector of a coherent state is

$$|\psi_c\rangle = \sum_{n=0}^{\infty} C_n^{coh} |n\rangle \quad (3)$$

Using Eqs.(1)-(3), one can show that for the case of multimode detection, the measured CF is $g^{(2)} = A + B/\langle n \rangle$ for the squeezed vacuum state, given by Eqs.(2a,b), where $A = 1$ and $B$ is the inverse number of detected frequency and spatial modes, while $g^{(2)} = 1$ for the coherent state, given by Eq.(3). The dependence of $g^{(2)}$ on the mean photon number for squeezed vacuum is a very typical feature, an indication of pairwise correlations. The significant advantage of $g^{(2)}$ measurement over a direct reconstruction of photon-number distribution is that it is insensitive to optical losses and finite efficiencies of the detectors, which may not be always precisely known and controlled in an experiment [27]. This makes $g^{(2)}$ measurements highly relevant in the study of single-photon sources, low-gain PDC and other applications.

Let us now derive an algorithm for obtaining $g^{(2)}$ from MPPC data. According to definition, see Eq.(1), for a given output of MPPC, one should define an equivalent number of double coincidences (numerator) and the squared total number of detected photons

(denominator). Under faint light pulses impinging on MPPC, it produces outputs of discrete amplitudes, equal to $kA_0$, where $k$ is the number of pixels having fired ($k=1,2,3...$) and $A_0$ is the amplitude of a standard pulse produced by an individual pixel. By recording results of repetitive measurements during the acquisition time, one obtains a histogram of MPPC counts $N_1, N_2...N_k$, where $N_k$ is the number of detected pulses with amplitudes equal to $kA_0$. The total number of registered photons is then $N^{(1)} = \sum_{k=1}^{\infty} k N_k$, and if the number of pixels is $m$, the average number of photons hitting a single pixel is $n^{(1)} = N^{(1)}/m$. Each pair of pixels represents a single HBT interferometer, the number of such interferometers can be estimated as $\binom{m}{2} = m(m-1)/2$, where $\binom{m}{2}$ is the number of 2-combinations in $m$ (the binomial coefficient). The total number of pairwise coincidences is $N^{(2)} = \sum_{k=2}^{\infty} \binom{k}{2} N_k$, and thus the number of pairwise coincidences per single HBT interferometer for large $m$ is $n^{(2)} \equiv 2N^{(2)}/m(m-1) \approx 2 \sum_{k=2}^{\infty} \binom{k}{2} N_k / m^2$. The resulting $g^{(2)}$ can be calculated as[1]

$$g^{(2)} = 2 \frac{\sum_{k=2}^{\infty} \binom{k}{2} N_k}{\left(\sum_{k=1}^{\infty} k N_k\right)^2}. \tag{4a}$$

Note that a similar algorithm can be used for the measurement of higher-order correlation functions given by Eq.(1):

$$g^{(l)} = \frac{m^l}{\binom{m}{l}} \frac{\sum_{k=l}^{\infty} \binom{k}{l} N_k}{\left(\sum_{k=1}^{\infty} k N_k\right)^l} \tag{4b}$$

Moreover, any spatially resolving single-photon detector can be used for these measurements, such as, single-photon sensitive charge coupling devices (EMCCD, ICCD) [28,29].

Let us now consider a realistic model of MPPC [23]. Since photon bunching of PDC is only pronounced at low photon numbers, we limit our approach by considering a simple linear model. A typical example of the histogram of MPPC outcomes along with the dark noise contribution is shown in Fig.1.

In order to account for the MPPC dark noise we estimated the probability of a dark noise event and a real photocount to appear simultaneously. It was found that the contribution of such a cross-term was almost negligible. Let the subscript "DN" denote quantities where the effect of the dark noise has been taken into account. We have

$$N_{k,DN} = N_k + n_k,$$

where $n_k$ is the number of occurrences of a $k$-photon event from the dark noise only, which can be measured independently.

---
[1] Note that Eq.(4) is independent of the detector quantum efficiency. To see this, one has to note that k is just a dummy variable for summing in Eq.(4). Hence, the effect of loss, which is to replace k with k' =k/η, where η is the quantum efficiency, will not affect the value of $g^{(2)}$.

Next, let us consider the effect of the crosstalk. In this case a $k$-photon event would contain a contribution from the lower photon number events with the addition from the crosstalk. Given the photon number distribution in Fig.1 and the crosstalk probability of about 7-15%, the gains and losses of more than one photon due to the crosstalk are ignored in the present model. Note that the dark noise is also affected by the crosstalk which in some cases predominates the generation of multiphoton dark noise events.

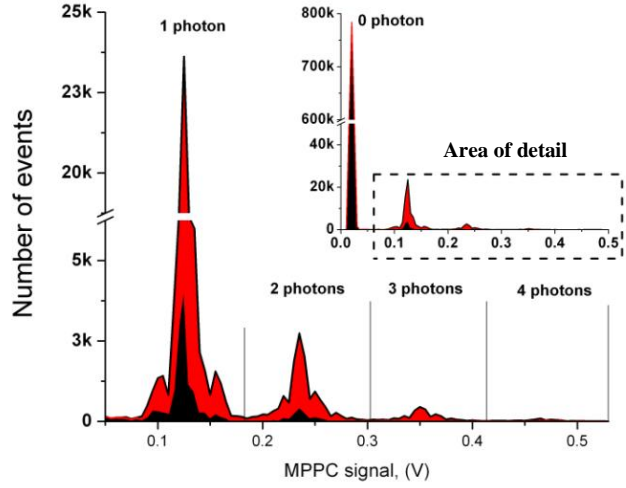

Fig.1 (Color online). An experimental histogram of the MMPC output for the signal mode of PDC (at 500nm) resulting in ~0.012 photons per pulse. The raw signal (red fill color) and the MPPC dark noise, which also includes the optical noise of the PDC setup (black fill color) were acquired for $10^6$ shots of the pump laser. Inset is the extended part of the histogram, which shows the vacuum component.

Let the subscript "CT" denote the quantities where the crosstalk has been taken into account and let $P$ be the probability of the occurrence of one crosstalk photocount. In this case, the number of $k$-photon events is given by

$$N_{k,CT,DN} = N_k + n_{k,CT} - kPN_k + (k-1)PN_{k-1}, \qquad (5)$$

The third term is the number of $k$-photon events that have been modified by the crosstalk and have become $(k+1)$-photon events, the fourth term is the number of $(k-1)$-photon events that have gained an extra photon from the crosstalk to become $k$-photon events.

Under the above assumptions, the dark count events add trivially to the photon statistics; hence, the expression in Eq.(5) can be simplified by subtracting the dark photocounts (with their cross-talk avalanches) and just consider the effect of the crosstalk:

$$N_{k,CT} = N_{k,CT,DN} - n_{k,CT},$$

The total number of photocounts is given in this case by

$$N_{Total,CT} = \sum_{k=1}^{\infty} k N_{k,CT} = (1+P) \sum_{k=1}^{\infty} k N_k \quad (6)$$

The measured $g^{(2)}$ function of the photocounts $N_{k,CT}$, according to Eqs.(4)-(6), becomes

$$g^{(2)} = \frac{(1+2P)}{(1+P)^2} g_0^{(2)} + \frac{2P}{(1+P)} \frac{1}{N_{Total,CT}}, \quad (7)$$

where $g_0^{(2)}$ is the correlation function of the incident light, which would be given by Eq.(4a) in the absence of the cross-talk. This result means that the counts of a detector with the crosstalk have the second-order normalized CF differing from the incident light CF by a normalization coefficient (very close to a unity) and an additive term scaling as $\propto 1/N_{Total,CT}$. The crosstalk probability $P$ can be easily measured using some light source with known $g_0^{(2)}$, for instance, coherent or multimode light with $g_0^{(2)} = 1$. Then, the unknown CF of the PDC source can be found by subtracting the additive contribution and normalizing the measured CF according to Eq.(7). This approach was realized in our experiment.

## 3. Experiment

In the experimental setup, shown in Fig.2, the 4-th harmonic of a pulsed Nd:YAG laser with the wavelength 266 nm (Crystalaser, 30 ns long pulses, 20kHz repetition rate) was used as a pump with its intensity controlled by a half wave plate and a polarizing beam splitter in the range from 20 uW to 5 mW. The long-term and the pulse-to-pulse stability of the UV-laser were controlled by a power meter and a fast Si photodiode (Thorlabs), respectively. The fluctuations of the UV-laser intensity did not exceed 3%. A lens with a focal length of f=750 mm was used to focus the pump beam into two 5mm long type-I BBO crystals, where PDC occurred. The crystals had oppositely directed axes in order to compensate for the spatial walk-off of the pump. After passing the BBO crystals the pump was eliminated by a UV mirror (UVM) whilst the PDC emission fluently passed through it. The collinear part of the PDC beam with the divergence equal to that of the pump was coupled into a single-mode optical fiber (SMF) by means of an achromatic lens f=6.2 mm, placed at a distance of 550 mm from the crystal. At the output of the fiber a collimated beam was formed by a lens with f=6.24 mm and addressed into an MPPC module (Hamamatsu, model C10751-02 with 400 pixels embedded in 1.5*1.5 mm chip). In order to ensure a relatively homogeneous illumination of the MPPC area, a lens (L) with f=400 mm was used to focus the beam onto the MPPC chip and provided a spot with the diameter 700 um. A fast digital oscilloscope (LeCroy Wave Runner, 2 GS/s sampling rate) was used to capture the analogue output from the MPPC (in a range 0-650 mV) and plot the histogram of its amplitude. The standard acquisition for each point was $10^6$ shots of the pump laser.

The results of the measurements on the coherent state were obtained with the attenuated continuous-wave Nd:YAG laser second-harmonic radiation with the wavelength 532 nm (Photoptech), whose intensity was modulated by an acousto-optic modulator (AOM, Gooch and Housego), driven by an external pulse generator. The laser pulse was tailored to mimic the profile and duration of the PDC pulse and its intensity was controlled by a half-wave plate (HWP) and a polarizing beam splitter (PBS). In relevant measurements the laser pulse was addressed into the optical path by a flipping mirror (FM1) and then was focused onto the MPPC chip with the same spot size as in the PDC experiment.

In order to compare the results of the MPPC measurements with a traditional HBT experiment, the beam was diverted by a flipping mirror (FM2) into an intensity interferometer with a 50/50 beam splitter (BS) and detected by two gated APDs (Perkin-Elmer SPCM AQR-14), whose outputs were addressed to a coincidence circuit with a 50 ns window. In order to avoid the APD saturation in experiments with high photon fluxes, the beam was attenuated by neutral density filters (not shown) to ensure that the counting rate of each APD was much lower than the repetition rate of the laser.

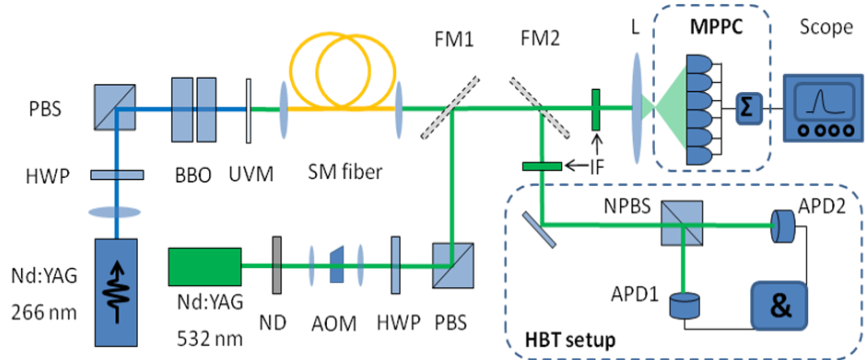

Fig. 2 Experimental setup (Color online). The 4-th harmonic of Nd:YAG laser at 266 nm is used to pump BBO crystals where PDC occurs. The pump intensity is controlled by a half wave plate (HWP) and a polarizing beam splitter (PBS). The collinear PDC is coupled into a single-mode fiber (SM fiber). In experiments with the coherent state a cw Nd:YAG laser at 532 nm is modulated by AOM and diverted by the flipping mirror (FM1). The flipping mirror (FM2) is used to address the beam to either the MPPC or the HBT scheme consisting of a nonpolarizing 50/50 beam splitter (NPBS), two APDs (APD1, APD2), and a coincidence circuit (&). UVM is a pump rejection mirror; IF are narrowband interference filters; ND, neutral density filter; L, a lens.

The dark noise of the MPPC was 11 dB suppressed by thermoelectrically cooling the chip down to -4.5C and sealing the detector in a custom-made housing with a coated optical window. The dark noise (which also accounts for the crosstalk) was about 25KCounts/sec and its gated part was carefully subtracted from the experimental data, as justified above. In the experiments with PDC the noise contribution also included the luminescence of the optical elements of the setup and it was measured by extinguishing PDC by tilting the polarization of the pump at 90 degrees by a half wave plate. Note that cooling of the chip did not help to suppress the crosstalk as confirmed by the results of other groups [30-32]. The MPPC output was synchronized with the laser sync pulse and was gated with 50 ns window.

## 4. Results and Discussion

First, $g^{(2)}$ measurements were carried out with the MPPC module by measuring the coherent state, given by Eq.(3), at different laser powers, which provided a reference with $g_0^{(2)} = 1$. For this purpose, a weak laser pulse was addressed into the measuring apparatus by erecting a flipping mirror FM1. Then the laser beam was gradually attenuated and a set of corresponding amplitude histograms was recorded. The dependence of $g^{(2)}$ versus the mean number of photocounts, calculated according to Eq.(4a) and fitted by Eq.(7) with $P$ being the only fitting parameter, is shown in Fig.3 and Fig.4 (black dashed trace, squares) and summarized in Table 1. The value of the crosstalk probability was found to be $P = 0,177 \pm 0.003$, with the coefficient of determination of the fit $COD(R^2) = 0,9801$. As shown in Fig.3, although excess two-photon correlations are not expected for the coherent state, the dependence of CF obtained by the MPPC exhibits strong photon bunching. This effect is attributed to the first-order crosstalk, which, according to Eq.(7), becomes especially pronounced at low photon numbers.

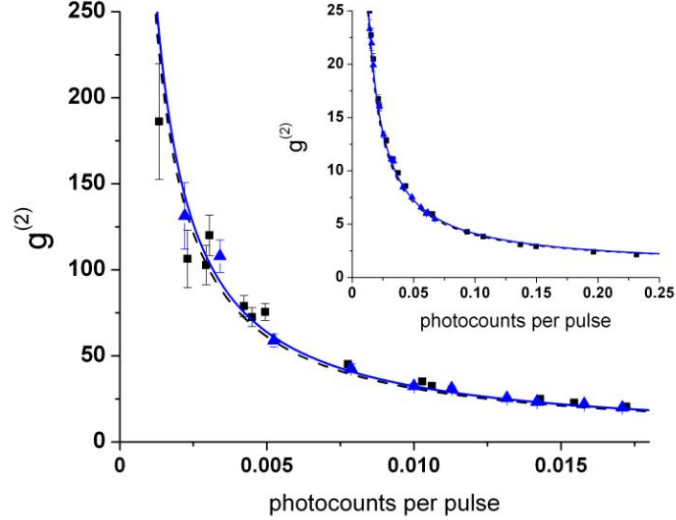

Fig.3 (Color online). Dependence of $g^{(2)}$ on the mean number of photocounts per pulse, obtained via MPPC measurements of the coherent state (black dashed trace, squares) and the signal radiation of the two mode squeezed vacuum state (blue solid trace, triangles). The curves are experimental fits with the corresponding parameters given in Table 1. Inset: the $g^{(2)}$ dependence at large numbers of photocounts.

Next, $g^{(2)}$ measurements were carried out by HBT and MPPC setups measuring only the signal radiation of two-mode squeezed vacuum, see Eq.(2a), which, in case of multimode detection also provides a reference with $g_0^{(2)} = 1$ [26,27]. For this purpose the FM1 was dissembled, so that the PDC beam from the SM fiber was addressed directly into the measuring apparatus. The optical axes of the crystals were set to 47.53 deg, which corresponded to the collinear generation of signal and idler photons at 500 nm and 568 nm, respectively. The signal photons fluently passed through a 10 nm FWHM interference filter, centered at 500 nm, while the correlated idler photon at 568 nm was discarded. The $g^{(2)}$ measured by HBT setup was found at a level of $1.10 \pm 0.1$ and demonstrated a flat dependence on the photon number, varied through changing the pump power. Then the PDC beam was addressed into the MPPC module and analogous to the measurements with the coherent state, a set of corresponding amplitude histograms was recorded. The dependence of $g^{(2)}(N_{Total,CT})$ calculated according to Eq.(4a) and fitted by Eq.(7) is shown in Fig.3 (blue solid trace, triangles) and summarized in Table 1. In order to compare the results with the coherent state, the photon numbers were adjusted for slight differences in quantum efficiency and transmission profile of the filter. The obtained value of the crosstalk parameter from the fit of the experimental data is $P = 0.186 \pm 0.002$ with $COD(R^2) = 0.991$. As shown in Fig.3, the obtained dependence demonstrates good agreement with the results obtained for the coherent state. Similar results have been obtained measuring the idler radiation at 568nm.

Table 1. Compilation of the fitting results [a] of the dependencies in Fig. 3

| | State and Measurement configuration | Figure and trace | $P$ | COD ($R^2$) |
|---|---|---|---|---|
| #1 | Coherent state measured with **MPPC** | Figures 3,4 **black dashed curve** | $0.177 \pm 0.003$ | 0.9801 |
| #2 | Signal radiation of two mode squeezed vacuum measured with **MPPC** | Figure 3 **blue solid curve** | $0.186 \pm 0.002$ | 0.991 |

[a] All the fittings were done using Origin software (OriginLab) by Levenberg-Marquardt algorithm, weighted for the experimental uncertainties.

Next, measurements of the single-mode squeezed vacuum state were made. The optical axes of the crystals were set to 47.63 deg which corresponded to the collinear frequency degenerate phase matching. An interference filter centered at 532nm with a 2 nm FWHM was used to ensure that both signal and idler modes were measured. Analogous to the previous case, the dependence of $g^{(2)}(N_{Total,CT})$ was calculated from the MPPC amplitude histograms according to Eq.(4a). The result is shown in Fig.4 (red solid trace, circles) and fitted by the parameters shown in Table 2. Note that the asymptotic value of the correlation function for the squeezed vacuum was taken from the results of the measurements of the coherent state. From the obtained results it is clearly seen that on top of the crosstalk, $g^{(2)}$ reveals additional two-photon correlations, which are attributed to the two-photon nature of squeezed vacuum.

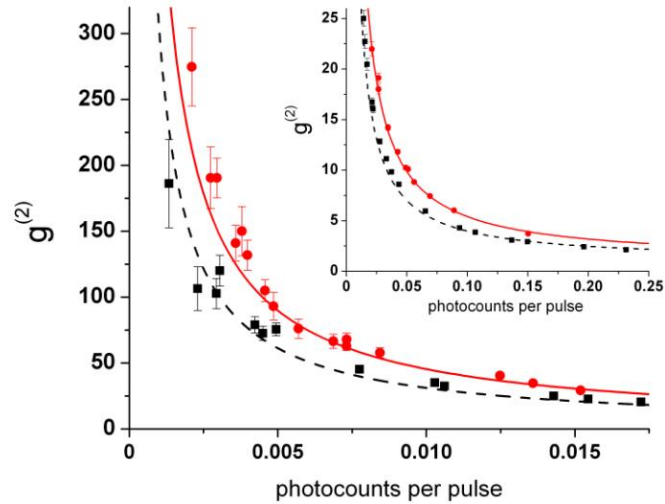

Fig.4 (Color online). Dependence of $g^{(2)}$ on the mean number of photocounts per pulse, obtained via MPPC measurements of the coherent state (black dashed trace, squares) and the single-mode squeezed vacuum state (red solid trace, circles). The curves are experimental fits with the corresponding parameters given in Table 1, 2. Inset: the $g^{(2)}$ dependence at large numbers of photocounts.

In order to infer the actual $g_0^{(2)}$ of the single mode squeezed vacuum from the MPPC measurements, Eq.(7) was used with $P = 0,177 \pm 0.003$ which was obtained from the earlier measurements of the coherent state. The resulting curve is shown in Fig.5 (red solid trace). The results are compared with the ones obtained in the HBT experiment, which are also shown in Fig.5 (black dashed trace, squares) and the fitting parameters are summarized in Table 2. The results of the measurements with MPPC and HBT setup demonstrate good agreement, whilst a slight discrepancy may be attributed to a higher-order crosstalk effect. Note that the mean photon numbers for both measurements shown in Fig.5 were adjusted by corresponding quantum efficiencies of the detectors and a crosstalk of MPPC. A direct comparison of single APD and MPPC module, realized in our setup by using attenuated laser pulses, revealed almost identical count rates of the detectors in the relevant dynamic range (average ratio 1.05). Taking into account the manufacturer's data on the quantum efficiency of the APD (equal to 50% at 532 nm) and the probability of the MPPC crosstalk, we found that the resulting quantum efficiency of the MPPC in our tests is equal to 41%. This figure represents a reasonable agreement with the value of 39% given in the datasheet of MPPC at -4.5C (which accounts for both the non-unity fill-factor of the chip and the quantum efficiency of individual pixels).

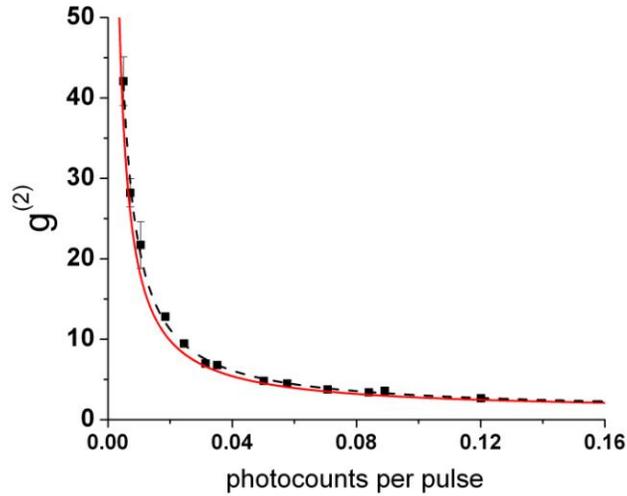

Fig.5 (Color online). Comparison of the dependence of the inferred $g^{(2)}$ on the mean number of photocounts per pulse for single-mode squeezed vacuum, measured by the MPPC (red solid trace), with the one obtained with a traditional HBT setup (black dashed trace, squares). The functions are presented in Table 2.

Table 2. Compilation of the fitting results of the dependencies in Figs. 4-5.

| | State and Measurement configuration | Figure and trace | $g^{(2)} = A + B/N_{total}$ | | COD ($R^2$) |
| --- | --- | --- | --- | --- | --- |
| | | | A | B | |
| 1 | Single-mode squeezed vacuum measured with MPPC | Figure 4 red solid curve | 0.977±0.01 | 0.444±0.004 | 0.991 |
| 2 | Inferred result for the single-mode squeezed vacuum measured with MPPC | Figure 5 red solid curve | 1±0.014 | 0.176±0.007 | (inferred) |
| 3 | Single mode squeezed vacuum measured with HBT | Figure 5 black dashed curve | 0.985±0.05 | 0.204±0.002 | 0.994 |

These results demonstrate that with a proper account for the crosstalk, MPPC detectors can be used for measurements of the second-order correlation function. Indeed, the measurement of the correlation function for some source with known statistics, such as, a coherent source, directly reveals the contribution of the crosstalk and allows one to subtract it.

Based on the present results and considering a remarkable capability of MPPC to distinguish higher photon numbers, our analysis can be further expanded to the studies of higher-order intensity correlation functions, see Eq.(4b). In such experiments the MPPC detector adjusted for the cross talk contribution would represent a compact and cost-effective solution in comparison with the existing detection techniques.

**5. Conclusion**

In conclusion, the second-order correlation function for coherent and squeezed vacuum states has been experimentally determined by using MPPC for the first time. We introduced an algorithm for obtaining the correlation function from any spatially resolved multi-pixel detector (MPPC, CCD). We developed a model of crosstalk in MPPC, which describes an additional term in the second-order correlation function. The validity of our approach was confirmed by demonstrating good agreement with independent measurements conducted in an HBT setup. It seems likely that at moderate gain of the OPO, when higher photon numbers are to be populated, the applicability of MPPC will be limited by the crosstalk. However, it is worth mentioning the continuing efforts on improving the fabrication process and design of APD arrays aimed to reduce the crosstalk and thus broaden the MPPC applicability [32]. Besides, the presented approach to correlation analysis can be directly applied to any kind of spatially resolving detectors such as, single photon resolving ICCD and EMCCD, where the influence of the crosstalk can be filtered out by addressing single pixels. Due to the fact that correlation measurements tolerate linear losses and finite efficiencies of the detectors, they are especially beneficial in experiments where the photon statistics have to be instantly estimated without knowing the exact nature of the losses. This approach may be exploited in a vast number of experiments, ranging from the studies of atomic-photon interfaces to astronomy applications.

**Acknowledgement**

We would like to thank Rodney Chua for the assistance in building the detector housing. This work was supported by A-STAR Investigatorship grant.